\documentclass[groupedaddress, twocolumn]{revtex4}

\usepackage{graphicx}   

\newcommand{\figref}[1]{\figurename~\ref{#1}}

\newcommand{\meter}[1][]{\ifx|#1|\unit{m}\else\unit[#1]{m}\fi}
\newcommand{\hertz}[1][]{\ifx|#1|\unit{Hz}\else\unit[#1]{Hz}\fi}
\newcommand{\fm}[1][]{\ifx|#1|\unit{fm}\else\unit[#1]{fm}\fi}
\newcommand{\fluence}[1][]{\ifx|#1|$\unit{mJ/cm^2}$\else$\unit[#1]{mJ/cm^2}$\fi}

\begin{document}

\title{Imaging the antiferromagnetic to ferromagnetic first order phase transition of FeRh}

\author{S. O. Mariager}
\email{simon.mariager@psi.ch} \affiliation{Swiss Light Source, Paul Scherrer
Institut, 5232 Villigen, Switzerland}
\author{L. Le Guyader}
\altaffiliation{Current address: Helmholtz-Zentrum Berlin fur Materialien und
Energie, Bessy II, 12489 Berlin, Germany} \affiliation{Swiss Light Source,
Paul Scherrer Institut, 5232 Villigen, Switzerland}
\author{M. Buzzi}
\affiliation{Swiss Light Source, Paul Scherrer Institut, 5232 Villigen,
Switzerland}
\author{G. Ingold}
\affiliation{Swiss Light Source, Paul Scherrer Institut, 5232 Villigen,
Switzerland}
\author{C. Quitmann}
\affiliation{Swiss Light Source, Paul Scherrer Institut, 5232 Villigen,
Switzerland}

\begin{abstract}
The antiferromagnetic (AFM) to ferromagnetic (FM) first order phase
transition of an epitaxial FeRh thin-film has been studied with x-ray
magnetic circular dichroism using photoemission electron microscopy. The FM
phase is magnetized in-plane due to shape anisotropy, but the
magnetocrystalline anisotropy is negligible and there is no preferred
in-plane magnetization direction. When heating through the AFM to FM phase
transition the nucleation of the FM phase occurs at many independent
nucleation sites with random domain orientation. The domains subsequently
align to form the final FM domain structure. We observe no pinning of the
FM domain structure.
\end{abstract}

\maketitle

\section{Introduction}
The metallic alloy FeRh undergoes an uncommon phase transition at T$_T\approx
105^o$C. Here the magnetic order upon heating changes from antiferromagnetic
(AFM) to ferromagnetic (FM) while the lattice simultaneously expands by $\sim
0.7~\%$.\cite{Fallot1939} As the transition is first order it involves a
latent heat and phase coexistence of the AFM and FM states. The value of
T$_T$ above room temperature is attractive for technological applications and
has lead to proposed uses in thermally assisted magnetic
recording\cite{Thiele2003}, magnetocaloric refrigiration\cite{Annaorazov1992}
and magnetostrictive transduction\cite{Ibarra1994}. For any application
utilizing the phase transition of FeRh an improved understanding of the
dynamics and the phase coexistence is of interest. The phase transition also
provides an opportunity to study the general phenomena of phase coexistence
in a first order magnetic phase transition. Other materials where a
magnetostructural transition plays a crucial role for the properties include
the paramagnetic to FM transition in MnAs \cite{Wilson1964, Kim2011} and the
interplay between magnetic and structural domains in shape memory alloys like
the Heusler alloy Ni$_2$MnGa.\cite{Ullakko1996} Phase coexistence also plays
a vital role in strongly correlated materials like Manganites.
\cite{Uehara1999,Dagotto2001}

\begin{figure} [b]
\includegraphics[scale=.5]{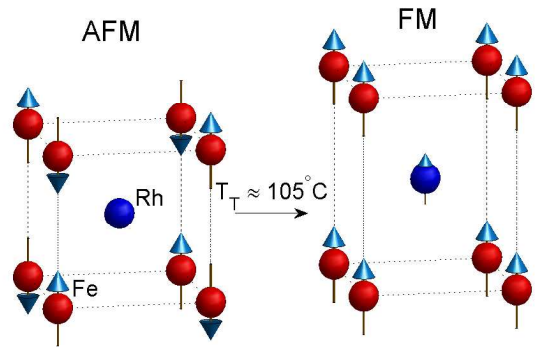}
\caption{(Color online) Sketch of the FeRh unit cell in the AFM and
FM phases. Red spheres symbolize Fe atoms and blue spheres Rh atoms
with the direction of magnetic moments indicated by the arrows. The
structural change is shown for the thin film case where expansion
only occur along the surface normal. The structural expansion has
been enhanced for clarity. } \label{figureSketch}
\end{figure}
FeRh has the CsCl crystal structure and the magnetic structure is known from
M\"{o}ssbauer spectroscopy\cite{Shirane1963}, neutron
diffraction\cite{Shirane1964} and band-structure calculations
\cite{Moruzzi1992}. The unit cell is sketched in \figref{figureSketch}. The
AFM phase is of type II with nearest neighbor Fe atoms aligned
antiferromagnetically with moments m$_{Fe} \approx 3~\mu_B$ and zero magnetic
moment on the Rh atoms. In the FM phase the Fe atoms align ferromagnetically
and a moment is induced on the Rh atoms, m$_{Rh} \approx 1~\mu_B$. In bulk
samples the magnetic transition is accompanied by an isotropic structural
expansion of $\sim 0.7~\%$, while for thin-films the in-plane expansion is
restricted by the substrate\cite{Maat2005}. Here the expansion at the phase
transition is only along the surface normal as sketched in
\figref{figureSketch}. While the electronic structure is largely unaffected
by the transition in both bulk crystals\cite{chen1988} and thin
films,\cite{Lee2010} the heat capacity, entropy and electrical resistance all
change at T$_T$\cite{Sharma2011, Annaorazov1996}. The exact transition
temperature depends on both composition\cite{McKinnon1970} and the addition
of transition metal impurities\cite{Kouvel1966}, which complicates comparison
between different experiments. Though the AFM to FM phase transition has been
known since 1939 \cite{Fallot1939} the physical mechanism behind it is still
debated. The first simple empirical model proposed historically assumed a
change of sign of the Fe-Fe interaction as a function of the lattice
expansion but failed to explain the experimental work, especially the large
change in entropy during the transition \cite{McKinnon1970}. On the other
hand density functional theory (DFT) computations reproduce the existence of
the AFM and FM phases and the volume dependency of the lattice constant
\cite{Moruzzi1992}. The low energy difference ($\sim$0.2~mRy/atom) between
the AFM and FM phases could be explained by considering the effect of spin
waves \cite{Gu2005} and recent local DFT work indicate that the volume
dependence of the AFM Fe-Fe exchange interaction combined with the unaffected
FM Fe-Rh interaction play a crucial role for the physical properties of FeRh.
\cite{Sandratskii2011}

The coexistence of the FM and AFM phases during the phase transition has been
reported in several recent experiments. X-ray magnetic circular dichroism
(XMCD) experiments were interpreted to conclude that even in the early stages
of the transition the microscopic FM domains were in the final FM state.
\cite{Stamm2008, Radu2010} The lattice expansion was studied by x-ray
diffraction which both directly showed the mixed phase and indicated that the
transition was initiated at the free surface of the sample.\cite{Kim2009}
Magnetic force microscopy images of polycrystalline samples showed a mixed
phase\cite{Yokoyama1998} and magnetization curves have been interpreted to
support an Avrami model behavior for the transition.\cite{Lu2010} Similarly,
when induced by a fs laser pulse we found that the transition proceeded
through the nucleation of many independent and initially unaligned FM
domains.\cite{Mariager2012} In order to obtain a microscopic image of the
coexisting magnetic phases a spatially resolved probe is however needed and
very recently \textcite{Baldasseroni2012} used XMCD photoelectron emission
microscopy (PEEM) to image the transition. This technique is well suited
because PEEM provides spatially resolved images with a resolution down to
50~nm, while XMCD gives excellent magnetic contrast in the FM phase.

In this paper we present XMCD PEEM images obtained at various temperatures of
an epitaxial FeRh film which was slowly heated and cooled through the AFM to
FM phase transition. We first analyze the domain structure and anisotropy of
the FM phase and then focus on images obtained during the phase transition
and show how the phase transition is initiated at many independent nucleation
sites. We present a quantitative analysis of the nucleation, and finally
compare the magnetic process to the structural change studied by x-ray
diffraction. This work complements both our previously published work on the
same sample on the ultrafast laser induced AFM to FM
transition\cite{Mariager2012} and the recent findings by
\textcite{Baldasseroni2012}.

\section{Experimental details}
\begin{figure} [t]
\includegraphics[scale=.65]{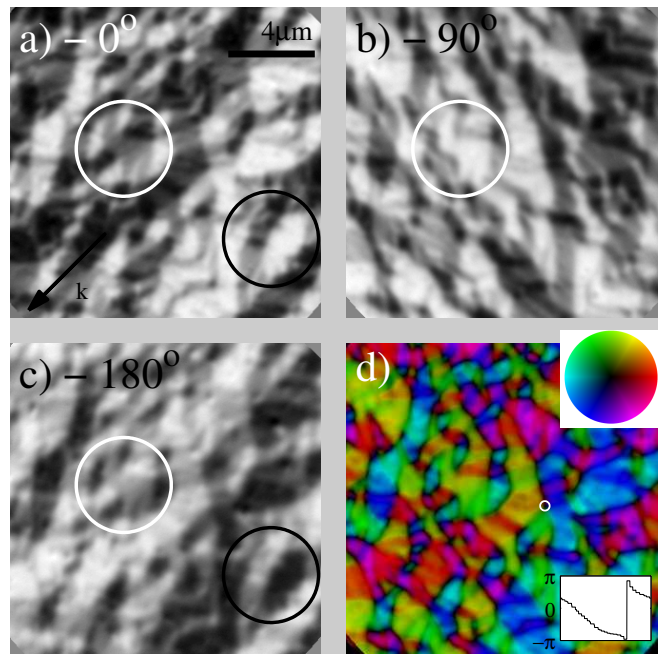}
\caption{XMCD PEEM images of the FeRh film taken at
the Fe L$_3$ edge at sample rotations of (a) 0$^o$, (b) 90$^o$ and
(c) 180$^o$ and T = 150$^o$C. The black circles highlight an example of
inverted contrast after 180$^o$ rotation. The white circles
highlight an example where the contrast change from grey to white to
grey. In (a) the field of view is indicated by the scale bar and the
direction of the x-rays is given by the arrow. This direction
roughly corresponds to FeRh [100]. (d) The derived in-plane domain
structure obtained from (a), (b) and (c). The color indicates the
in-plane azimuth angle $\phi$ of the magnetization, while the
saturation of the color is proportional to the magnitude as shown by the color
wheel. The lower insert shows the azimuthal angle $\phi$ along the contour
corresponding to the white circle.} \label{figure1a}
\end{figure}
The FeRh thin film (\textit{d} = 47 nm) was grown on MgO (001) by
co-magnetron sputtering from elemental targets \cite{Maat2005}. The single
crystal film was epitaxial to the substrate with a (001) surface and
$[100]_{FeRh}\parallel [110]_{MgO}$ as characterized by x-ray diffraction.

The XMCD PEEM measurements were done at the SIM beamline of the Swiss Light
Source\cite{Flechsig2010}. This beamline provides circularly polarized light
with an energy resolution  of E/$\Delta E \approx 5000$. The x-ray incidence
angle was 16$^o$ and all images were recorded at the Fe L$_{3}$ edge at
708~eV. The field of view was varied from 10-20$~\mu$m and images were
recorded with 512$\times$512 pixels and a spatial resolution of $\sim$ 50~nm.
The electron microscope accelerates and detects the emitted photoelectrons
and the depth of view is around 5~nm, limited by the mean free path of the
photoelectrons. The XMCD images are obtained from two images taken with left
(-) and right (+) circular polarization and x-ray intensity I. The asymmetry
is calculated for each pixel as:
\begin{equation}
I_{XMCD} = (I_+-I_-)/(I_++I_-) \propto \textbf{K}\cdot\textbf{m}
\label{eqn1}
\end{equation}
The asymmetry is proportional to the magnetization \textbf{m} projected onto
the x-ray wavevector \textbf{K}.\cite{Scholl2002} In this case all
non-magnetic contrast mechanisms cancel because they are independent of the
photon helicity. By measuring $I_{XMCD}$ for three different sample
orientations it is possible to obtain three projections of \textbf{m} to
determine its 3 vector components.\cite{Guyader2012} The sample was heated by
a small resistive heater, with rates of 0.85$^o$~K/min during heating and
-0.7$^o$~K/min during cooling.

The x-ray diffraction (XRD) experiments were performed at the MicroXAS
beamline of the Swiss Light Source. We used a 7~keV x-ray beam in a gracing
incidence geometry with incidence angle 0.71$^{\circ}$ which matches the
x-ray penetration depth to the film thickness. The (101) Bragg reflection was
measured with a PILATUS 100K pixel detector in a rocking curve scan where the
sample was rotated around the surface normal in 60 discrete steps in the
interval $\pm2^o$. By recording the data in a 1D scan with a 2D detector the
Bragg reflection can be mapped in 3D, which allows tracking of both in-plane
and out-of-plane peak shifts and hence the shift in lattice
constant\cite{Mariager2009}. The XRD measurements were performed in air and
the temperature was changed in discrete steps while allowing the temperature
to stabilize between each measurement.

\section{Ferromagnetic Domain Structure}
\begin{figure} [t]
\includegraphics[scale=1]{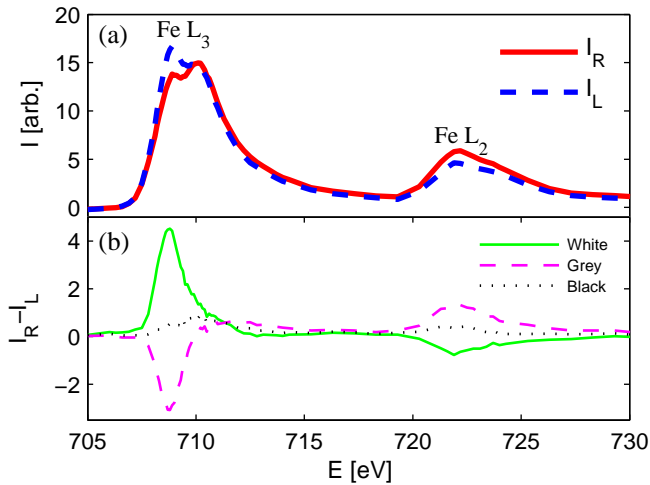}
\caption{(Color online) (a) Intensity of the emitted photo electrons
as a function of x-ray energy for left and right circularly
polarized light. The Fe L$_{3}$ and L$_{2}$ edges are visible, and
both exhibit a splitting. (b) XMCD for three
different domains. The dashed line (magenta) corresponds to the
spectra in (a).} \label{figure2}
\end{figure}
In order to analyze the domain structure of FeRh it is necessary to obtain
XMCD PEEM images at least at three different rotations of the sample, and in
\figref{figure1a} we show three PEEM images obtained at $0^o$, $90^o$ and
$180^o$, with the rotation taken around the surface normal. The sample
temperature was 150$^o$C, well above T$_T$. In all the images FM domains are
clearly visible as black and white contrast and the lengthscale of the domain
structures is on the order of $1~\mu$m, as seen by comparison to the 4$~\mu$m
scalebar. The net moment in the images is essentially zero, as expected when
no external field was applied to the sample.

The rotation of the sample by 180$^o$ from \figref{figure1a}a to c results in
a reversal of contrast from black to white and vice versa, as highlighted by
the black circles and evident throughout the images. This indicates in-plane
magnetic domains because out-of-plane domains do not change contrast upon
rotation of the sample. Consistent with this, a 90$^o$ rotation results in an
interchange of grey domains (\textbf{m} $\perp$ \textbf{H}) with white/black
domains. This is highlighted for a single feature by the white circles in
\figref{figure1a}a-c. The images thus directly shows that the FeRh film is
dominated by in-plane magnetization. This magnetic structure is due to the
shape anisotropy of the thin film, where the in-plane magnetization minimizes
the stray field and therefore the total energy.
\begin{figure} [t]
\includegraphics[scale=1]{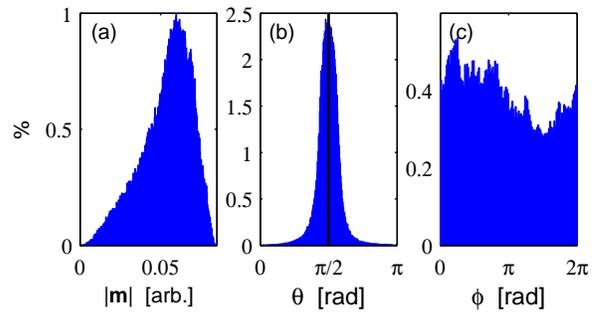}
\caption{Histograms obtained from the analysis of the local magnetic
moment in \figref{figure1a}d. (a) the magnitude of the magnetic
moment, (b) the out-of-plane angle $\theta$ and (c) the azimuth
angle $\phi$.} \label{figure1b}
\end{figure}
In \figref{figure2}a we show the absorption spectrum from a single domain for
respectively left and right circular polarized x-rays. Both the Fe L$_{3}$
edge at 708~eV and the L$_{2}$ edge at 721~eV are split in two peaks of which
only the peaks at lower energy show dichroism. This is a typical sign of
surface oxidation \cite{Regan2001}. In this experiment the sample surface was
neither capped nor sputtered prior to imaging which explains the surface
oxidation. We did subsequently confirm that sputtering 2~nm off the surface
removed the oxide layer on a second sample (not shown). In \figref{figure2}b
we show the asymmetry for three different domains, corresponding to a
respectively white, black and grey area in \figref{figure1a}a. This confirms
that the patterns in \figref{figure1a}a do indeed arise from the magnetic
structure of the sample.

\begin{figure*} [t]
\includegraphics[scale=.9]{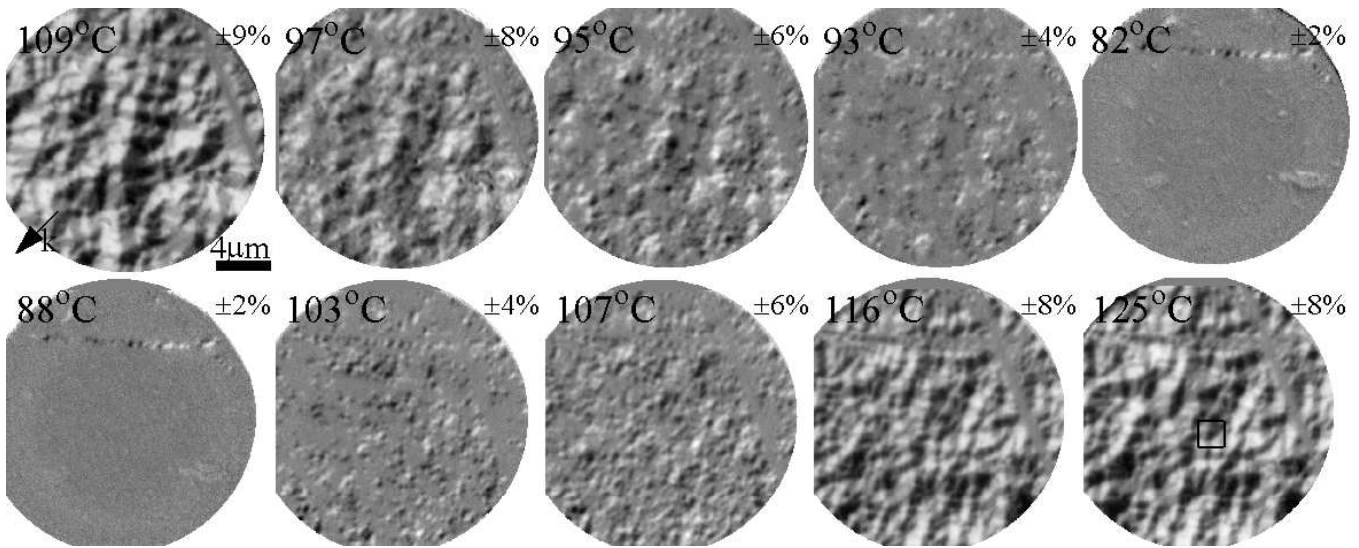}
\caption{XMCD PEEM images obtained during cooling (upper row) and
heating (lower row) through the AFM to FM phase transition in FeRh
and labeled by the temperature. The field of view is 20$~\mu$m as
indicated on the scale-bar in the first image. The arrow shows the
direction of the x-ray wavevector. In order to maximize the contrast
the gray-scale is not the same for all images. The number in the
upper right corner of each image gives the limits on the interval
used for the gray scale. The first image uses a gray scale running
from -9\% (black) to 9\% (white). The black square in the last image
shows the region which is magnified in \figref{figure3b}. }
\label{figure3}
\end{figure*}

The three images in \figref{figure1a} were used to determine the
magnetization \textbf{m(r)}.\cite{Guyader2012} In \figref{figure1a}d we show
the azimuth angle of the in-plane magnetization, where the color saturation
is a quantitative scale for the magnitude of the moment. \figref{figure1a}d
reveals domains with a typical lengthscale of 1$~\mu$m and fairly straight
domain walls. It also shows the existence of nodes around which the in-plane
magnetization rotates 360$^o$. One example is shown in the lower insert in
the figure. Here the azimuthal angle is taken along the contour shown as a
white circle in \figref{figure1a}d, and the 0 to 360$^o$ degree rotation is
evident. As a consequence there must be an out-of-plane divergence but these
vortex cores are too small to be resolved. Previously reported domain
patterns for FeRh polycrystalline bulk samples\cite{Yokoyama1998} were likely
caused by dipolar interaction between the grains, a feature which is not
present in our epitaxial film.

The reconstruction of the three dimensional magnetization vector allows us to
determine if the film has preferred directions of the local magnetic moments.
In \figref{figure1b} we show histograms of the three spherical coordinates of
\textbf{m}: the moment magnitude $|\mathbf{m}|$, the out-of-plane inclination
angle $\theta$ and in-plane azimuth angle $\phi$. Here $\theta$ lies in the
range from 0 to $\pi$ with 0 corresponding to the magnetic moment lying along
the surface normal and $\pi/2$ being in-plane. The magnitude of the moment
shows one major peak, but is significantly broadened towards zero, because we
do not have the resolution to image domain walls. Due to this areas of the
sample where the direction of the local magnetic moment changes on a scale
shorter than the PEEM resolution will then be measured as having a smaller
magnitude of the moment. The histogram of the out-of-plane angle $\theta$ has
a clear peak at $\pi/2$, confirming the in-plane magnetization which is also
visible from the raw images in \figref{figure1a}. For the azimuth angle
$\phi$ we find a fairly uniform distribution while a strong anisotropy in the
plane would give well defined domains and reveal itself as distinct peaks in
the histogram. The absence of such peaks indicates that FeRh has a very low
magnetocrystalline anisotropy. We are not aware of a measurement of the
anisotropy constant K for FeRh, but our result supports the assumption that K
is small for FeRh.\cite{Guslienko2004,Bordel2012}

From the presence of two peaks in a histogram of the asymmetry of a single
XMCD image \textcite{Baldasseroni2012} concluded that their FeRh film had a
four fold anisotropy. However, since the asymmetry for an in-plane magnetized
surface is proportional to $\cos(\phi)$ an isotropic distribution of azimuth
angles also results in two peaks in the histogram of the asymmetry. Based on
our azimuthal study of the magnetization vector we conclude in contrast to
ref. 28 that the crystalline anisotropy is too weak to give a well defined 4
fold easy axis in FeRh.

It has also been suggested that the tetragonal distortion imposed by the
substrate leads to a magnetocrystalline anisotropy favoring out-of-plane
magnetization of the FM phase, when the ratio between the lattice constants
satisfy $c/a > 1$ as for our sample.\cite{Bordel2012} The shape anisotropy
was however estimated to be several times larger than the magnetocrystalline
anisotropy for a 150~nm film and our results clearly show that the film is
100\% in-plane magnetized. This shows that the proposed magnetocrystalline
anisotropy if present is too weak to overcome to the shape anisotropy of our
50~nm thin film.

\section{Magnetic Phase Transition}
Having established the ability to map FM domain structures we now discuss the
AFM to FM phase transition and the phase coexistence. In \figref{figure3} we
show images taken as the sample was cooled (upper row) and heated (lower row)
through the AFM to FM transition. The images in \figref{figure3} show a
different region of the sample than the images in \figref{figure1a}. The
feature consisting of two straight lines which cross in the upper right
corner of each image appears to be a scratch on the sample surface, and was
used to monitor the drift of the sample. In the following we refer to each
image in \figref{figure3} by the temperature at which it was obtained.

\begin{figure} [t]
\includegraphics[scale=1]{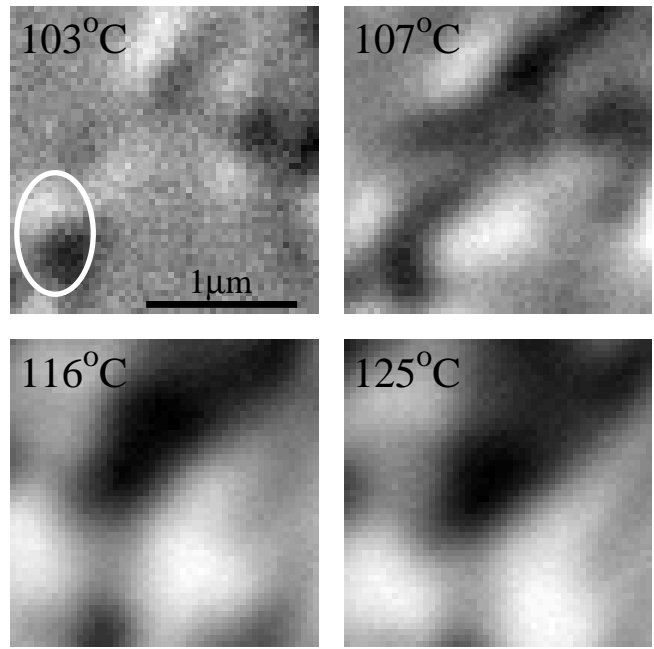}
\caption{XMCD-PEEM images obtained during heating. The images show a
magnified region of the images in the lower row of \figref{figure3}.
The white circle highlights a small FM domain with an apparently
closed magnetization loop.} \label{figure3b}
\end{figure}

We first focus on the heating process. In the AFM phase (T = 82$^o$C and T =
88$^o$C) no domain structure is visible, because the AFM phase does not show
any XMCD. The growth of the FM phase in the AFM matrix then proceeds through
the nucleation of many independent domains on a sub micron length-scale (T =
103$^o$C). This directly shows the coexistence of the two magnetic phases
expected for a first order phase transition at intermediate temperatures. As
the temperature is increased (T = 107$^o$C) the dominant feature remains
nucleation of new domains rather than growth of previously nucleated domains.
Like the final domain pattern in \figref{figure1a} the initially nucleated
domains have no preferred in-plane magnetization directions. As the film
reaches a purely FM state neighboring domains to some extent realign upon
contact (T = 116$^o$C). This is evident in \figref{figure3b} where the
initial black contrast of the circled feature changes to white. The result is
in the final structure which as already observed is dominated by domains of
size $\sim1~\mu m$ (T = 125$^o$C). Once the transition reaches this state the
domain structure is unchanged upon further heating. One example of this
nucleation process is highlighted in \figref{figure3b} which shows a
magnification of a $2\times2~\mu m^2$ area corresponding to the square in
\figref{figure3} (T = 125$^o$C). At T = 103$^o$C several independent FM
domains are visible and in several cases positive (white) and negative
(black) dichroism are paired. This indicates that the independent FM domains
form closed magnetic loops in order to minimize the stray field. As the
temperature is raised these smaller structures change or grow to larger
domains. \figref{figure3} and \figref{figure3b} show how the AFM to FM
transition proceeds through the nucleation of many independent unaligned
domains, and subsequent partial reorientation of neighboring domains. This
observation agrees with the recent work by \textcite{Baldasseroni2012} and
with our previous results where we have found the same mechanism to dominate
the phase transition when it is induced with a fs laser
pulse.\cite{Mariager2012} We find the same nucleation dynamics on both ps and
s timescales, suggesting that also the laser induced phase transition is a
thermal process.

Because we can only image the FM domains the observed process during cooling
is slightly different from heating. In this case we observe how the FM
domains disappear, rather than how the AFM phase nucleates in the FM matrix.
In \figref{figure3} we see how the FM domains upon cooling first start to
shrink (T = 97$^o$~C) and eventually break up (T = 95$^o$~C) into independent
domains. The result is again a process where the final coexistence of the AFM
and FM phase consists of many small sub-micron size FM domains (T = 93$^o$~C)
in an AFM matrix. This process is similar to heating, but there is less
realignment of the FM domains. To quantify this difference we define
realignment as a change in sign of dichroism and determine the number of
pixels which change sign at some point during heating (cooling). In this case
the fraction of the film which realign is 14\% during heating and 7\% during
cooling. We stress that this definition of realignment is a simplification
and that the actual fraction of the film which realigns is higher.

The images in \figref{figure3} taken at T = 109$^o$C and T = 125$^o$C were
obtained before and after cooling to the AFM phase. When comparing the two
images the two domain structures appear to be independent. To quantify this
we again sort pixels according to whether they show positive or negative
dichroism, and compare how many pixels are positive in both images. We find a
fraction of 0.26, which is practically identical to the 0.25 expected if
there was no correlation between the domains in the two images.

A comparison of images during the late stages of cooling (T = 93$^o$~C) and
early stages of heating (T = 103$^o$~C) however reveal a statistical
significant correlation of the FM regions \footnote{In both of the images
$\approx19\%$ of the sample is FM, while 7\% of the sample is FM in both
images. This is more than 3 standard deviations higher than the overlap found
in a numerical simulation where a 19\% FM area was chosen randomly in two
images and then compared.}. That is, on average the regions which nucleate
first during heating also retain the FM states longest during cooling. The
magnetization direction is however not correlated in the two images which
again shows that the film has no memory of the previous FM phase. This rules
out the existence of a single unique pathway between the AFM and FM domain
patterns.

\begin{figure} [t]
\includegraphics[scale=1]{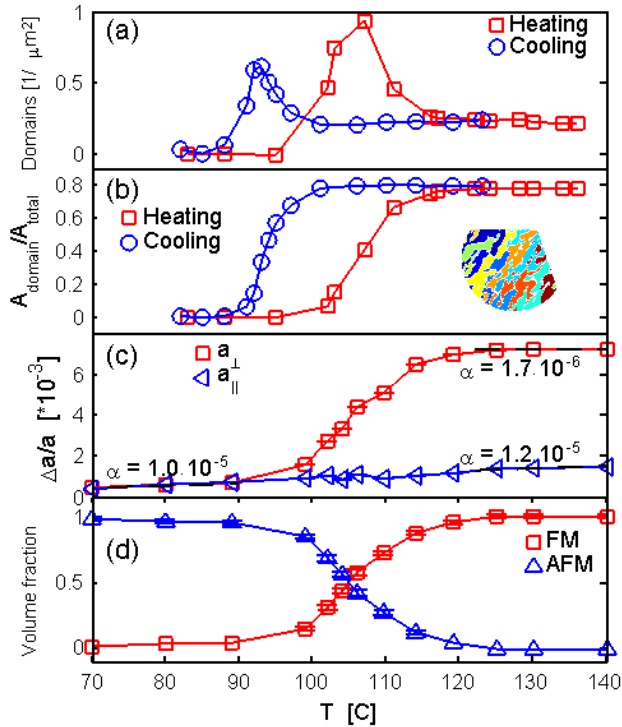}
\caption{(Color online)  (a) Domains per $\mu m^2$ during cooling
(blue circles) and heating (red squares). A domain is defined as a
connected area with either positive or negative dichroism. (b) The
total area of the FM domains. The insert show the domain structure
corresponding to \figref{figure3} T = 125C. (c) In-plane (blue
triangles) and out-of-plane (red squares) expansion of the lattice.
(d) Volume fractions of the FM (red squares) and AFM (blue
triangles) phases obtained during heating.} \label{figure4}
\end{figure}

To obtain a more quantitative understanding of the phase transition the
images in \figref{figure3} were analyzed as follows. Due to the low
magnetocrystalline anisotropy it is not possible to identify domains by the
strength of the asymmetry. In addition the XMCD contrast is significantly
weaker at the early stages of the nucleation, which complicates the
comparison of domains at different temperatures. Instead we define a domain
as a connected area of pixels with either positive or negative dichroism.
This is not a precise definition because it depends upon sample orientation
(\figref{figure1a}), but it allows us to identify the appearing FM nuclei and
quantify their number and sizes as well as the total FM area. The result of
this analysis is shown in \figref{figure4}. \figref{figure4}a shows the
number of individual domains as a function of temperature for both heating
and cooling, while \figref{figure4}b shows the FM fraction of the film. The
final FM fraction only reaches 0.8 because we do not measure domains with a
magnetic moment perpendicular to the x-ray beam which have no magnetic
contrast. In reality the entire film is in the FM phase. The quantitative
analysis presented in \figref{figure4} support the qualitative interpretation
of the images in \figref{figure3}. The peak seen at intermediate temperatures
in \figref{figure4}a confirms that the nucleation proceeds through an
initially large number of domains which later realign resulting in fewer
domains. We note that the definition of a domain used in this analysis will
underestimate the final number of domains according to the standard
definition.

We finally note that while a FM surface layer has previously been reported
below T$_T$ in the AFM phase for both Au, Al and MgO capped FeRh films
\cite{Ding2008,Baldasseroni2012,Fan2010} we clearly observe no such FM
surface layer for the uncapped sample studied here. Though we will not
discuss the subject further, we have observed FM surface layers on some
uncapped films and imaged FeRh films with 2~nm Pt cap layer without a FM
surface. The many different results thus seems to indicate that details in
film growth and stoichiometry plays a bigger role in defining the magnetic
surface properties than the choice of capping.

\section{Structural Phase Transition}

The AFM to FM phase transition in FeRh has two connected components as the
change in magnetic order goes hand in hand with a structural expansion. To
complete the picture of the transition we show in \figref{figure4}c the
lattice expansion as a function of temperature. The shift in lattice constant
$\Delta$a can to first approximation be found from the shift in Bragg peak
position $\Delta$q because $\Delta a/a \approx - \Delta q/q$. In addition to
the thermal expansion, the out-of-plane lattice constant shows a clear shift
of 0.7~\% at T$_T$. The in-plane lattice constant only shows linear thermal
expansion through the entire temperature range. The in-plane thermal
expansion $\alpha_{\parallel} \approx 1.1\cdot10^{-5}$ of the FeRh film is
close to the value for MgO ($\alpha_{MgO} = 1.04\cdot10^{-5}$ at T =
300~K).\cite{Browder1969} The one dimensional expansion thus appears to be
due to in-plane pinning to the MgO substrate. This difference from the
isotropic expansion in bulk crystals has been reported
previously.\cite{Maat2005}

In the diffraction experiment we measure the total diffraction from the two
coexisting phases. The Bragg peak measured during the phase-transition is a
superposition of two peaks originating from the AFM and FM phases
respectively and can be decomposed into two peaks corresponding to the two
phases\cite{Mariager2012}. As the integrated intensity is proportional to the
scattering volume the volume fraction of the AFM and FM phase as a function
of temperature can then be determined. In \figref{figure4}d we show the
volume fractions of the AFM and FM phase during the transition. We find
agreement between the evolution of the FM structure and the magnetic order.
The small differences in temperature is expected given the difference in how
the temperature was changed and measured.

\section{Discussion}
The high resolution magnetic images obtained of the phase transition and the
phase coexistence in FeRh allows us to speculate about which microscopic
mechanisms drive the phase transition. Here three observations play a crucial
role. First, the many independent nucleation sites present on a sub-micron
length scale indicate that the driving mechanism must appear on a similar or
smaller length scale. Second, the correlation between images taken during
cooling and heating show that there exist either pinning defects or local
impurity fluctuations leading to local variations in transition temperature.
Third, the magnetization direction in the FM phase is not pinned neither for
the final structure nor during nucleation.

Concerning the existence of pinning centres our spatial resolution makes it
difficult to discern between a transition that appears at topographic
features such as defects or strain fields as suggested for the AFM to FM
transition in Gd$_5$Ge$_4$ \cite{Moore2006} or due to a broader but fixed
local variation of the transition temperature such as suggested for Ru doped
CeFe$_2$ \cite{Roy2004} and the Heusler alloy NiMnIn.\cite{Sharma2010} Clear
topographic features such as the scratches visible in \figref{figure3} or
intentionally produced antidots (not shown) however do not dominate as
nucleation centres. In addition rather than observing nucleation only at
fixed sites we observe nucleation in all regions but at different
temperatures. These two observations suggest that local impurity variations
lead to changes in the volume free energy large enough to compensate for the
interfaces created during phase coexistence.\cite{Imry1979} The resulting
heterogenous nucleation was also observed in previous work on FeRh.
\cite{Maat2005,Lu2010,Baldasseroni2012}

The many independent nucleation sites could support the recent theoretical
suggestion that local fluctuations in magnetic moment and/or volume acts as
driving "forces" for the transition \cite{Derlet2012}. The many independent
nucleation sites then indicate a high attempt frequency and/or a low energy
barrier for the transition. This can also explain the fact that the domain
pattern is dominated by many small domains because the existence of many
independent nucleation centers means that even a moderate domain wall
coercivity will result in many small magnetic domains.

We did not apply a magnetic field to the sample because an external magnetic
field distorts the path of the photoelectrons and hereby reduces the
resolution of the microscope. In an applied magnetic field it could be
studied if the nucleated FM phase nucleate aligned to the field, align to the
field immediately after nucleation or only after a full FM domain pattern is
achieved. We have previously indirectly observed the latter case at fields of
0.1~T when the phase transition was induced by a fs laser\cite{Mariager2012}.
This question of alignment relates to the question of how the AFM and FM
domains are related to one another, if at all. A theoretical pathway has been
used in calculations, where the Fe moments rotate $90^o$ during the
transformation from AFM to FM \cite{Sandratskii2011}. In this case an AFM
domain can give rise to four different directions of the magnetic moment in
the FM phase. From our comparison of images taken at the late stages of
cooling and early stages of heating, we confirm that there is no unique
pathway between the AFM and FM order. It was also calculated and shown for
one film that the tetragonal crystal structure imposed on the film by the
substrate leads to a magnetocrystalline anisotropy which strengthens this
$90^o$ rotation of the moments.\cite{Bordel2012} Depending on the preferred
magnetization direction this effect might however be negated by the shape
anisotropy, as in our experiment. To verify a possible relation between the
AFM and FM domain structures an experimental probe of the AFM domain
structure is needed. These are scarce but one candidate could be x-ray linear
magnetic dichroism\cite{Spanke1998,Stohr1999}. While this technique has been
applied successfully to oxides only few studies of metallic systems exist.

\section{Conclusions}
To summarize we used XMCD PEEM to measure the FM domain structure and XRD to
confirm the structural changes of an epitaxial FeRh thinfilm. The film was
in-plane magnetized due to shape anisotropy and with no preferred in-plane
direction of the magnetization, which indicates a low magnetocrystalline
anisotropy. The resulting domain structure had a lengthscale of $1~\mu m$ and
the magnetic domains were not pinned when the film was repeatedly cycled
through the transition. We directly observed the phase coexistence expected
for a first order phase transition in the XMCD PEEM images, which are ideally
suited to study this phenomenon. We found that the phase transition proceeds
by nucleation at many small and independent sub-micron sized sites. This
effect dominates over growth of already existing FM domains and neighboring
domains only subsequently and to some extent realign to form the final FM
domain pattern. These findings match our previously developed model for the
laser induced phase transition\cite{Mariager2012} and the existence of the
same dynamics on ps and s timescale suggest that the laser induced phase
transition is also a thermal process. A further understanding of the AFM to
FM pathway and the underlying driving forces could be obtained by imaging
also the AFM phase or by time-resolved XMCD PEEM.

\begin{acknowledgments}
The PEEM images were obtained at the X11MA beamline, and the x-ray
diffraction experiments were performed on the X05LA beam line, both at the
Swiss Light Source, Paul Scherrer Institut, Villigen, Switzerland. We thank
D. Grolimund and C. Borca of X05LA for help and P. Derlet for fruitful
discussions. We thank E. E. Fullerton of U. C. San Diego for preparing the
sample. This work was supported by the Swiss National Foundation through NCCR
MUST.
\end{acknowledgments}


\end{document}